\documentclass[12pt,preprint]{aastex}
\shorttitle{SMA Observations of R\,CrA}%
\shortauthors{Chen \& Arce}%

\begin{document}

\title{R\,CrA SMM\,1A: Fragmentation in A Prestellar Core}

\author{Xuepeng~Chen\altaffilmark{1}, \& H\'{e}ctor~G.~Arce\altaffilmark{1}}

\affil{$^1$Department of Astronomy, Yale University, Box 208101, New Haven, CT 06520-8101, USA}

\begin{abstract}

We report the discovery of multiple condensations in the prestellar core candidate SMM\,1A 
in the R\,CrA cloud, which may represent the earliest phase of core fragmentation observed 
thus far. The separation between the condensations is between 1000 and 2100\,AU, and 
their masses range from about 0.1 to 0.2\,$M_{\sun}$. We find that the three condensations 
have extremely low bolometric  luminosities ($< 0.1$\,$L_{\sun}$) and temperatures ($< 20$\,K), 
indicating that these are young sources that have yet to form protostars. We suggest that these 
sources were formed through the fragmentation of an elongated prestellar core. Our results, 
in concert with other observed protostellar binary systems with separations in the scale of 
1000\,AU, support the scenario that prompt  fragmentation in the isothermal collapse phase 
is an efficient mechanism for wide binary star formation, while the fragmentation in the 
subsequent adiabatic phase may be an additional mechanism for close ($\leq$\,100\,AU) 
binary star formation.

\end{abstract}


\section{INTRODUCTION}

The origin of binary stars is still a puzzle in our understanding of star formation. 
Numerous theoretical simulations support the hypothesis that the fragmentation 
of collapsing molecular cloud cores, promoted by either rotation or turbulence, 
is the main mechanism for the formation of binary/multiple stellar systems  (see 
reviews by Bodenheimer et al. 2000, Tohline 2002, and Goodwin et al. 2007). 
Nevertheless, many key questions concerning this fragmentation process, e.g., 
when exactly does the fragmentation start, are still under debate (see Tohline
2002). Although it is generally assumed that cloud cores do not fragment during 
the free-fall collapse phase, several groups (e.g., Bate \& Burkert 1997; Tsuribe 
\& Inutsuka 1999; Boss et al. 2000) found that fragmentation can occur at the 
end of the isothermal phase (generally referred to as prompt fragmentation), while 
others (e.g., Truelove et al. 1997, 1998; Machida et al. 2005, 2008) argue that 
the isothermal gas is stable against fragmentation and found that fragmentation
only occurs during/after the adiabatic phase (see Figure~9 in Andr\'{e} et al. 2009 
for a discussion of the different evolutionary phases of core collapse).

On the observational side, a handful of young protostellar (i.e., Class\,0) binary 
systems have been found (e.g., Looney et al. 2000; Launhardt 2004), and there 
are increasing kinematics studies of binarity in the protostellar phase (e.g., Chen 
et al. 2007, 2008, 2009; Volgenau et al. 2006; Maury et al. 2010). However, the 
number of observed and well-studied protostellar binary (protobinary) systems is 
still small, and these systems have already gone through the fragmentation process. 
The observational link between the initial conditions in a fragmenting cloud core and 
the final stellar systems formed therein is still missing. It is therefore critical to observe, 
at high angular resolution, more dense cores in nearby molecular clouds to search for 
the earliest phase of fragmentation and study in detail their properties, in order to put 
direct constraints on fragmentation models.

In this Letter, we present Submillimeter Array\footnote{The Submillimeter Array is a 
joint project between the Smithsonian Astrophysical Observatory and the Academia 
Sinica Institute of Astronomy and Astrophysics and is funded by the Smithsonian 
Institution and the Academia Sinica.} (SMA; Ho et al. 2004)  dust continuum 
observations toward the R~Corona~Australis (R\,CrA) region. At a distance of
$\sim$\,170\,pc (Knude \& H{\o}g 1998), the CrA dark cloud is one of the nearest 
star-forming regions. As the most active star formation site in this cloud, the R\,CrA 
region has been extensively observed in the past two decades (see review by 
Neuh$\ddot{\rm a}$user \& Forbrich 2008, and references therein). Using SCUBA on 
the James Clerk Maxwell Telescope (JCMT), Nutter et  al. (2005) found a prestellar core 
candidate termed SMM\,1A in the R\,CrA region, which is elongated in the east-west 
direction and has a gas mass of $\sim$\,10\,$M_\odot$, and an effective radius of 
$\sim$\,3000\,AU. With a maximum velocity dispersion of about 0.8\,km\,s$^{-1}$ 
(Harju et al.~1993), the SMM\,1A core is gravitationally bound. Infall motions in this 
region of the cloud further confirm that this is a prestellar core (Groppi et al. 2004). 
In this Letter, we report the discovery of a multiple system within the SMM\,1A core, 
based on high angular resolution millimeter observations. This may represent the 
earliest phase of core fragmentation observed thus far.

\section{OBSERVATIONS AND DATA REDUCTION}

The R\,CrA region was observed with the SMA on 2006 August 20 in the compact 
configuration. Six antennas were used in the array, providing baselines from 5\,k$\lambda$ 
to 52\,k$\lambda$ at 220\,GHz. The SMA primary beam is about 55$''$ at this frequency. 
Two positions next to each other at 
(R.A.,~decl.)$_{\rm J2000}$\,=(\,19:01:53.3, $-$36:57:21.0) and 
(R.A.,~decl.)$_{\rm J2000}$\,=(\,19:01:56.4, $-$36:57:27.0) were observed. 
The digital correlator was set up to cover the frequency ranging 216.6$-$218.8\,GHz 
and 226.6$-$228.8\,GHz in the lower and upper sidebands, respectively. 
The 1.3\,mm dust continuum emission was recorded with a total bandwidth of 
$\sim$\,3.3\,GHz ($\sim$\,1.8\,GHz USB and $\sim$\,1.5\,GHz LSB). System 
temperatures for R\,CrA ranged from 100 to 280\,K (depending on elevation), with 
a typical value of $\sim$\,200\,K.

The visibility data were calibrated with the MIR package (Qi 2005), with quasars
3c454.3 and 1924-292 as the bandpass and gain calibrators, respectively. Uranus 
was used for absolute flux calibration, from which we estimate a flux accuracy of 
$\sim$\,20\%, by comparing the measured quasar fluxes with those listed in the
SMA calibration database. The calibrated visibility data were imaged using the 
Miriad toolbox (Sault et al. 1995). The SMA synthesized beam size at 1.3\,mm 
dust continuum, with robust {\it uv} weighting 0, is 5.7$''$\,$\times$\,2.3$''$.

\section{RESULTS}

Figure~1a shows the SMA 1.3\,mm dust continuum image of R\,CrA, overlaid  with the SCUBA 
850\,$\mu$m dust continuum contours (from Groppi et al. 2007). The northern part of this image
shows clear detection of the dust continuum emission associated with the Herbig-Ae star R\,CrA 
and protostellar cores SMM\,2, SMM\,1B and SMM\,1C --- these last two are also referred to as 
SMA1 and SMA2 in Groppi et al. (2007; see also Choi et  al. 2008). To the south, the SCUBA 
850\,$\mu$m image shows the SMM\,1A core, which is elongated in the east-west direction (see
Fig.\,1a). At higher angular resolution, the SCUBA 450\,$\mu$m image in Figure~1b shows that
the SMM\,1A core is clearly resolved into two sub-cores (see also van~den~Ancker~1999). At
even higher resolution, the SMA 1.3\,mm continuum observations reveal that the SMM\,1A
core is divided into three condensations, aligning roughly from east  to west, which we refer to
here as SMM\,1A-a, SMM\,1A-b, and SMM\,1A-c (Fig.\,1b). All are detected with signal-to-noise
ratio of 4 or more. Source SMM\,1A-a coincides with the eastern sub-core observed in the SCUBA 
450\,$\mu$m image, while SMM\,1A-b and SMM\,1A-c are coincident with the western sub-core 
(Fig.\,1b). The single-dish and interferometer observations are in general agreement with each 
other, which indicates that the three continuum sources detected with the SMA are neither artifacts 
nor are due to  noise in the interferometer image\footnote{The three sources in SMM\,1A were not
detected in the SMA 1.1\,mm observations in Groppi et al. (2007), probably due to the reasons: (1) 
the three sources lie near the edge of their primary beam ($\sim$\,46$''$ at 270\,GHz); (2) more flux
was resolved out in their extended configuration observations; and (3) the 1.1\,mm observations 
lack the sensitivity to detect these weak sources.}. Also note that the SiO\,(5--4) line and several other 
molecular lines, e.g., DCN\,(3--2), were included in this SMA correlator setup, but no line emission 
was detected from the three condensations.

Figure~2 shows the mid-infrared images of the R\,CrA region, taken by the {\it Spitzer Space 
Telescope}. The MIPS\,1 band image at 24\,$\mu$m (Fig.\,2a) clearly shows infrared emission 
at the position of cores SMM\,1B (associated with the infrared source IRS\,7B, see Wilking et al. 
1997), SMM\,2, and the Herbig stars R\,CrA and T\,CrA. There is also 24\,$\mu$m emission from the 
SMM\,1C core, which may be associated with the radio source B9 centered on the core (see Choi 
et al. 2008), and the infrared source IRS\,7A located $\sim$\,5$''$ to the south. The MIPS\,2 band 
image at 70\,$\mu$m (Fig.\,2b) has a lower angular resolution and is dominated by the emission 
from the SMM\,1C/IRS\,7A region. The MIPS\,3 (160\,$\mu$m) image of R\,CrA is extremely 
saturated, and is thus not shown here. None of the $Spitzer$ images at bands from 3.6 to 
70\,$\mu$m shows compact infrared emission in the SMM\,1A core, which suggests that the core 
is extremely cold. 

From the SMA 1.3\,mm images, we derived the positions, fluxes, and (deconvolved) FWHM sizes 
of the three sources in R\,CrA SMM\,1A using a multi-component Gaussian fitting routine (see 
Table~1). Angular separations between sources SMM\,1A-a and -b and SMM\,1A-b and -c are 
12\farcs2\,$\pm$\,0\farcs3 and 5\farcs8\,$\pm$\,0\farcs3, corresponding to projected separations 
of $\sim$\,2100\,AU and $\sim$\,1000\,AU (at a distance of 170\,pc), respectively.
Assuming that the 1.3\,mm dust continuum emission is optically thin, the total gas mass 
($M_{\rm gas}$) of the three sources is calculated with the same method as described 
in Launhardt \& Henning (1997). In the calculations, we adopt a dust opacity of 
$\kappa_{\rm d}\,=\,0.5\,{\rm cm}^2\,{\rm g}^{-1}$ (Ossenkopf \& Henning 1994), a typical 
value for dense and cold molecular cloud cores, and a mass-averaged dust temperature 
of $\sim$\,18\,K (see below). The gas masses of the three sources, derived from the 
SMA 1.3\,mm dust continuum observations, range from $\sim$\,0.10 to $\sim$\,0.23\,$M_\odot$ 
(see Table~1). 
With these mass and the FWHM size of the condensations, we estimate their 
average densities to be in the order of $10^7$ to $10^8$\,cm$^{-3}$. Note that these are 
average densities and the local peak densities in these sources may be orders of magnitude 
higher. These high densities, as well as the infall motions detected in this region (see Groppi 
et al. 2004), suggest that the three condensations in the SMM\,1A are on their way to form 
low-mass stars. Considering that the SMM\,1A core has a mass of approximately 10\,$M_\sun$,
and assuming a core-to-star efficiency of $\sim 30$\% (Evans et al.  2009), it is probable that a 
multiple stellar system with a total mass of 3 $M_\sun$ will eventually form in SMM\,1A.

\section{DISCUSSION}

\subsection{The Earliest Phase of Binary/Multiple Star Formation}

Table~1 lists the 450\,$\mu$m and 850\,$\mu$m fluxes estimated from the SCUBA images, 
and the 800\,$\mu$m fluxes adopted from van~den~Ancker (1999). Since sources SMM\,1A-b 
and SMM\,1A-c are not resolved in the JCMT observations, we treat them as one sub-core 
here. The SCUBA 450\,$\mu$m fluxes are derived by fitting two-dimensional Gaussians 
toward the two sub-cores (SMM\,1A-a and SMM\,1A-b+c), while the 850\,$\mu$m fluxes 
are estimated from the fluxes enclosed within roughly one beam around each sub-core.  
Based on the SMA 1.3\,mm and JCMT submm data points, as well as the 3\,$\sigma$ upper 
limits in the $Spitzer$ MIPS images, we construct the spectral energy distributions (SEDs) of 
the two sub-cores (see Figure~3). In order to derive luminosities and temperatures, we first 
interpolate and then integrate the SEDs, always assuming spherical symmetry. Interpolation 
between the flux densities is done by a $\chi$$^2$ single-temperature grey-body fitting to all 
points (including the upper limits). The resulting bolometric luminosities are 
$<$\,0.08\,$L_\odot$ for SMM\,1A-a and $<$\,0.09\,$L_\odot$ for SMM\,1A-b+c. The dust and 
bolometric temperatures are between 17 and 19\,K for the sources.

The low luminosities and temperatures, as well as the fact that no compact infrared 
emission is detected from these sources in the $Spitzer$ images, resemble the typical
properties of prestellar cores (see review by Andr\'{e} et al. 2009). In the VLA survey 
of the R\,CrA region, no radio source is detected at the positions of the three condensations
presented here (Choi et al. 2008). In contrast, the protostars SMM\,1B (Class\,0/I) and 
SMM\,1C (Class\,0) to the north of SMM\,1A are bright and centrally peaked at millimeter 
wavelengths (Fig.\,1a),  and are associated with hard X-ray (Forbrich \& Preibisch 2007), 
infrared (Fig.\,2a), and centimeter radio counterparts (Choi et al. 2008). Therefore, although 
the chemical and  kinematic properties of SMM\,1A are still poorly known, the observations 
thus far suggest that  the three sources in SMM\,1A are in an earlier evolutionary stage than 
that of Class 0 protostars. As predicted by theoretical studies, there are different phases in 
the evolution from an initial isothermal (10\,K) prestellar core to a Class 0 protostar (Andr\'{e}  
et al. 2009). The low luminosities ($<$\,0.1\,$L_{\sun}$) and temperatures (close to 20\,K) 
of the three sources in SMM\,1A are similar  to the predicted properties of so-called first 
hydrostatic (adiabatic) cores, which are formed after the isothermal collapse of prestellar
cores (e.g., Masunaga et al. 1998; Chen et al. 2010). However, it must be noted that large 
uncertainties remain in our estimates due to the limited observations available. Further high 
angular resolution observations at different wavelengths are needed to constrain the SEDs 
of the three sources in order to address more precisely their evolutionary statuses. Regardless 
of this, it is fair  to say that {\it the R\,CrA SMM\,1A multiple system represents the earliest phase 
of core fragmentation observed thus far}.

\subsection{Prompt (Isothermal) Fragmentation vs. Adiabatic Fragmentation}

Prompt fragmentation of rotationally flattened cloud cores with initially flat density profiles, 
immediately after a phase of free-fall collapse, has long been suggested as one of the 
most efficient mechanisms for binary star formation (Tohline 2002). Since this fragmentation 
is expected to occur at the end of the isothermal collapse phase, it is also referred to as 
isothermal fragmentation. The separation of the fragments formed via prompt fragmentation 
has a scale of 10$^2$$-$10$^4$\,AU, which corresponds to the Jeans length in this phase. 
In contrast, other groups found that isothermal gas is stable against fragmentation and that core 
fragmentation can only occur during (and after) the adiabatic phase (e.g., Truelove et al. 1998; 
Machida et al. 2008). Because this fragmentation occurs after the formation of the first adiabatic 
core, it is also called the first core fragmentation (Machida et al. 2005). Here, the separation of 
the fragments formed generally has a scale of 3$-$300\,AU (Machida et al. 2008). Clearly, 
observing the earliest phase of core fragmentation is the most effective way to discriminate 
between prompt (isothermal) and adiabatic fragmentation scenarios. If prompt fragmentation 
does occur in the isothermal phase, we may expect to find a binary/multiple system consisting 
of adiabatic cores with the separation of $\sim$\,1000\,AU.

As shown in the SCUBA images (see Figure~1b), the R\,CrA SMM\,1A core is highly elongated 
with an aspect ration of roughly 3, showing a morphology similar to the bars or filaments seen 
in the simulations of prompt (isothermal) fragmentation (see, e.g., Bate \& Burkert 1997; Tsuribe 
\& Inutsuka 1999; Boss et al. 2000). From the mass  and radius of the SMM\,1A core (see Nutter 
et al. 2005), the average volume density is calculated to be 1$-$2\,$\times$\,10$^{7}$\,cm$^{-3}$. 
The corresponding Jeans length is $\sim$\,1000\,AU for a temperature of $\sim$\,10\,K. The 
separations among the three sources found in the SMM\,1A core are also consistent with this 
estimated Jeans length. Moreover, our results from the SED fitting (see above) suggest that the 
three sources are at an evolutionary stage earlier than that of Class\,0 protostars, perhaps in the 
first (adiabatic) core stage. All these results appear to support the view that the three sources are 
formed through the prompt fragmentation of an elongated collapsing core during the isothermal 
phase. In this scenario, we would expect the rotational axis of SMM\,1A to be perpendicular to the 
direction of  the core's major axis, which would be roughly perpendicular to the large scale rotation 
observed in the R\,CrA region by Groppi et al. (2004). This is not necessarily inconsistent with our 
picture as other cores show drastic differences in the rotation axes measured at different scales 
(e.g., Ohashi et al. 1997). We also note that so far most observed protostellar binary/multiple 
systems (e.g., L\,723, Launhardt 2004; or NGC\,2264 D-MM1, Teixeira et al. 2007) have separations 
in the scale of 1000\,AU, which also supports the scenario of prompt fragmentation, although these 
protostellar systems have had at least a few 10$^4$ yrs to evolve from their prestellar stage and 
thus their separations may not represent real initial conditions.

On the other hand, as introduced above, prompt (isothermal) fragmentation has trouble accounting 
for the close binary systems with separations $\leq$\,100\,AU. The fragmentation of cores at the 
adiabatic phase is still an attractive mechanism for close binary star formation. Considering that the 
binary separation distribution in (pre-)\,main sequence stars peaks at $\sim$\,30\,AU, we actually would
expect to find frequently close protobinary systems with separations of $\leq$\,100\,AU to support the adiabatic 
fragmentation scenario. In practice, however, the number of this kind of systems is very small. This 
observational result can be understood by the fact that current large millimeter interferometers (e.g., SMA 
and IRAM-PdBI) normally reach 1--2$''$ angular resolution under general conditions, and thus mostly 
resolve wide protobinary systems with separations of 200\,AU or more in nearby star-forming clouds.
We believe that routine observations at angular resolution better than 0.5$''$ will reveal more close 
protobinary systems, which will then allow us to study in detail a statistically significant number of 
protostellar binary/multipe systems with a wide range of separations (from $\sim$\,10\,AU to 10$^4$\,AU).

\section{SUMMARY} 

We present SMA 1.3\,mm dust continuum observations toward the R\,CrA region. 
The 1.3\,mm dust continuum emission is detected from dense cores SMM\,1B,  
SMM\,1C, and SMM\,2, and the Herbig-Ae star R\,CrA in this region. We discover 
within the prestellar core candidate SMM\,1A a multiple system made up of three 
condensations with masses in the range of 0.1 to 0.2\,$M_{\sun}$. The angular 
separations between the three new sources are $\sim$\,12.2$''$ and $\sim$\,5.8$''$, 
corresponding to the projected separations of $\sim$\,2100\,AU and $\sim$\,1000\,AU, 
respectively.  Based on SMA observations and complementary data from the JCMT 
and $Spitzer$ telescopes, we construct their spectral energy distributions (SEDs) 
and find that all three sources have extremely low bolometric luminosities ($<$\,0.1\,$L_{\sun}$) 
and temperatures ($<$\,20\,K). We suggest that the three condensations in SMM\,1A 
resulted from the fragmentation of an elongated core in the isothermal phase. The 
R\,CrA SMM\,1A system appear to be the earliest phase of low-mass core 
fragmentation observed thus far.


\clearpage

\clearpage


\begin{deluxetable}{lllcccccc}
\tabletypesize{\scriptsize} \tablecaption{SMA 1.3\,mm and JCMT submm dust continuum results of R\,CrA SMM\,1A.\label{tbl-1}} \tablewidth{0pt}
\tablehead{\colhead{Source} & \colhead{R.A.$^{a}$} & \colhead{Dec.$^{a}$} & \colhead{$S_{\nu}$(1.3\,mm)$^{a}$} & \colhead{FWHM$^{a}$} & \colhead{$M_{\rm gas}$$^b$} & \colhead{$S_{\nu}$(450\,$\mu$m)}&\colhead{$S_{\nu}$(800\,$\mu$m)}&\colhead{$S_{\nu}$(850\,$\mu$m)}\\
\colhead{} & \colhead{(J2000)} & \colhead{(J2000)} & \colhead{[mJy]} & \colhead{maj.$\times$min.} & \colhead{[$M_{\odot}$]} & \colhead{[Jy]} & \colhead{[Jy]} & \colhead{[Jy]}}

\startdata
SMM\,1A-a          & 19:01:55.86 & $-$36:57:48.8  & 46$\pm$9      & 5\farcs6$\times$5\farcs2   & 0.10$\pm$0.02  & 5.0$\pm$2.0 & 1.5 &1.0$\pm$0.2\\
SMM\,1A-b          & 19:01:54.87 & $-$36:57:45.2  & 64$\pm$13    & 6\farcs8$\times$6\farcs3   & 0.13$\pm$0.02  & 4.6$\pm$1.8$^c$  & 0.8$^c$ & 0.7$\pm$0.2$^c$\\
SMM\,1A-c          & 19:01:54.56 & $-$36:57:40.6  & 110$\pm$20  & 7\farcs8$\times$3\farcs1   & 0.23$\pm$0.05  &  4.6$\pm$1.8$^c$  & 0.8$^c$ & 0.7$\pm$0.2$^c$\\
\enddata
\tablenotetext{a}{Center position, flux, and FWHM size of the
SMA 1.3\,mm dust continuum sources derived from the Gaussian fitting.}
\tablenotetext{b}{Total gas mass derived from the SMA 1.3\,mm dust continuum observations (see text
for the dust opacity and temperature used).}
\tablenotetext{c}{Combined fluxes for both sources SMM1A-b and SMM1A-c.}
\end{deluxetable}

\clearpage

\begin{figure*}[hlpt]
\begin{center}
\includegraphics[width=10cm,angle=0]{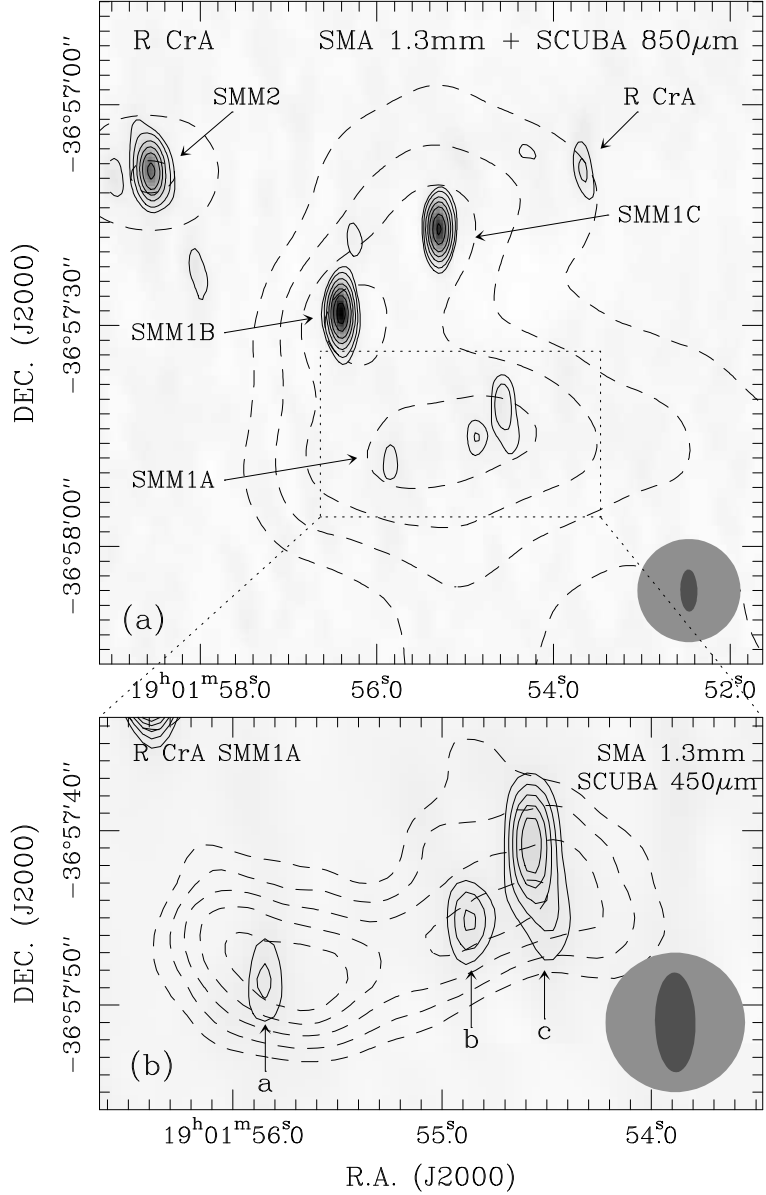}
\caption{\footnotesize (a)  SMA 1.3\,mm continuum data (solid contours) overlaid on the JCMT/SCUBA 
850\,\micron\ continuum map (dashed contours) of the R\,CrA region. The SMA data contours correspond 
to 3, 5, 9, 14, 20\,$\sigma$ levels and then increase in steps of 10\,$\sigma$ 
(1\,$\sigma$\,$\sim$\,6.2\,mJy\,beam$^{-1}$). The SCUBA 850\,\micron\ contours begin at 30\% and then 
increase in steps of 20\% of the peak emission ($\sim$\,2.4\,Jy\,beam$^{-1}$). The synthesized SMA beam 
(5.7$''$\,$\times$\,2.3$''$) and the HPBW of the SCUBA ($\sim$\,14$''$) are shown as grey ovals in the 
bottom right corner. (b) SMA 1.3\,mm dust continuum image (solid contours) overlaid on the JCMT/SCUBA 
450\,\micron\ continuum map (dashed  contours). Here, the SMA data start at the 3\,$\sigma$ level and 
increase in steps of 1\,$\sigma$. The SCUBA 450\,\micron\ contours (dashed lines) begin at 75\% and then 
increase by a step of 5\% of the peak emission ($\sim$\,10.2\,Jy\,beam$^{-1}$). The synthesized SMA beam 
and the HPBW of the SCUBA map ($\sim$\,8$''$) are shown in the bottom right corner.
 \label{rcra_sma_scuba}}
\end{center}
\end{figure*}

\clearpage
\begin{figure*}[hlpt]
\begin{center}
\includegraphics[width=16.5cm,angle=0]{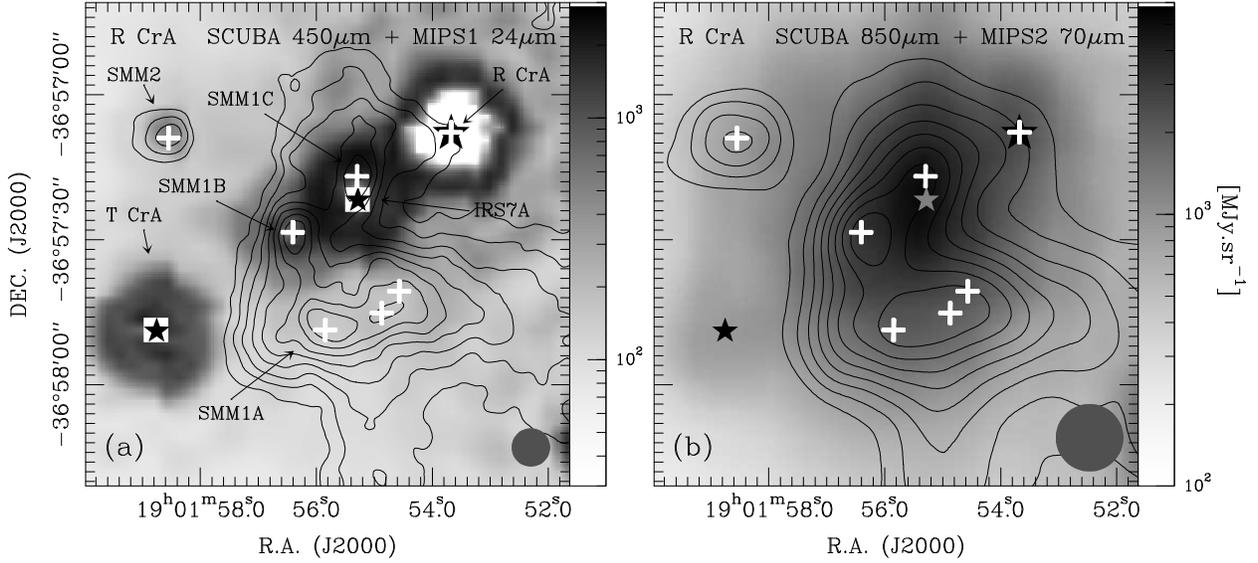}
\caption{(a) $Spitzer$ MIPS\,1 (24\,\micron)\ image of the R\,CrA region,  with contours showing the SCUBA
450\,\micron continuum emission. Contours levels begin at 20\% and increase by steps of 10\% of 
the peak emission ($\sim$\,10.2\,Jy\,beam$^{-1}$). (b) $Spitzer$ MIPS\,2 (70\,\micron)\ image of
the R\,CrA region, with contours showing the SCUBA 850\,\micron\ continuum emission. Contour 
levels begin at 20\% and increase in steps of 10\% of the peak emission ($\sim$\,2.4\,Jy\,beam$^{-1}$).
In both images the crosses mark the positions of the SMA 1.3\,mm dust  continuum sources,
while star symbols show the positions of the Herbig stars R\,CrA and T\,CrA, and the infrared source 
IRS\,7A (the MIPS\,1 image is saturated at the position of these three sources).
\label{rcra_spitzer_scuba}}
\end{center}
\end{figure*}

\clearpage
\begin{figure*}[hlpt]
\begin{center}
\includegraphics[width=12cm,angle=0]{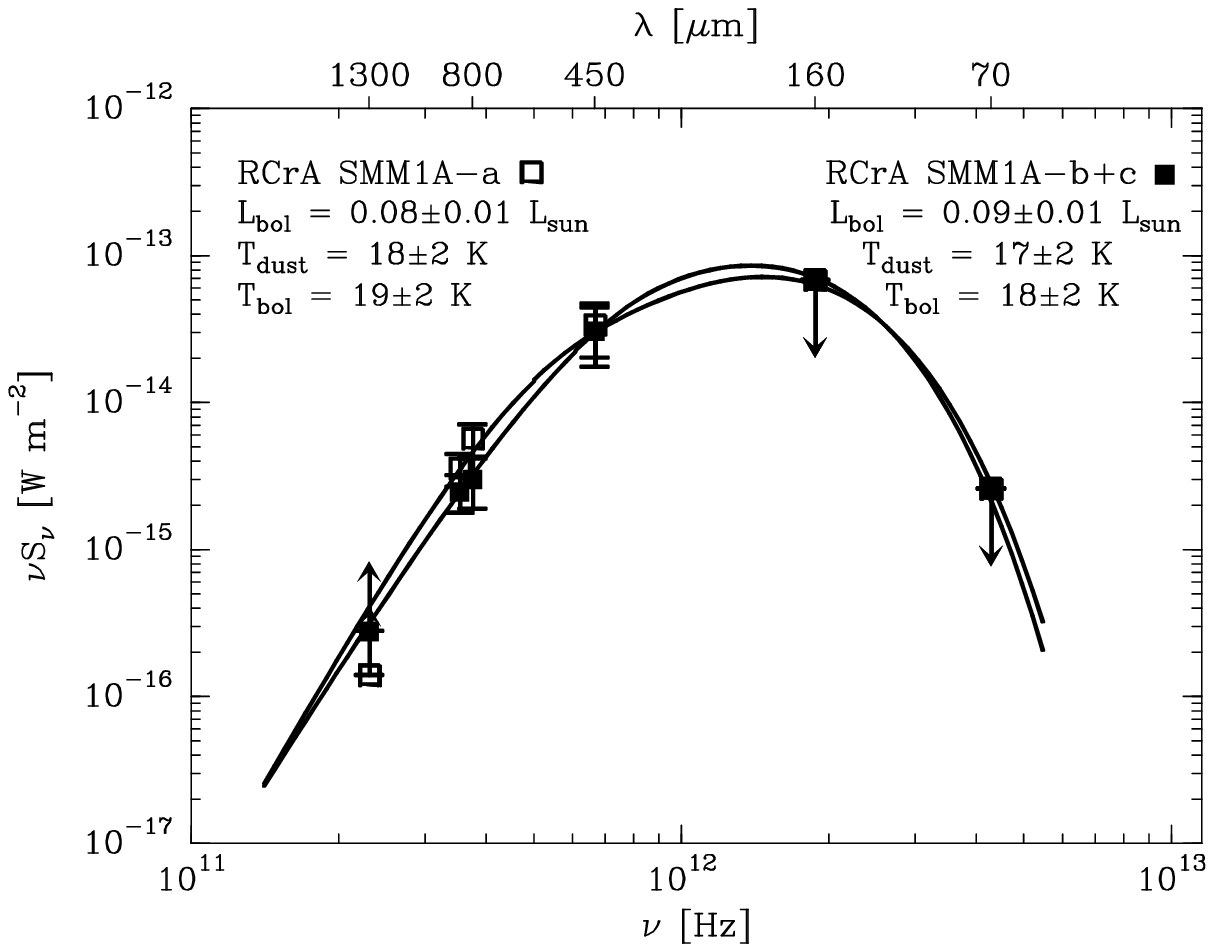}
\caption{Spectral energy distributions of the R\,CrA SMM\,1A-a and the SMM\,1A-b+c sub-cores. 
The $Spitzer$ 70\,$\mu$m and 160\,$\mu$m points, and the SMA 1.3\,mm point represent upper and lower 
limits, respectively (indicated by the arrow). Solid lines show the best-fit for all points at 
$\lambda$ $\geq$ 70\,$\mu$m using a greybody model 
$S_{\nu}$\,=\,$B_{\nu}$($T_{\rm d}$)(1\,$-$\,$e$$^{-\tau_\nu}$)$\Omega$, 
where $B_{\nu}$($T_{\rm d}$) is the Planck function at frequency $\nu$ and dust
temperature $T_{\rm d}$, $\tau_\nu$ is the dust optical depth as a function of 
frequency, and $\Omega$ is the solid angle of the source. The fitting results are 
summarized in the plot.\label{rcra_sed}}
\end{center}
\end{figure*}


\begin{thebibliography}{}

\bibitem[Andre et al.(2009)]{andre09} Andr\'{e}, P., Basu, S., \&
Inutsuka, S.-I. 2009, in Structure Formation in Astrophysics, ed.
G. Chabrier (Cambridge: Cambridge University Press), 254

\bibitem[Bate \& Burkert(1997)]{bb97} Bate, M.~R., \& Burkert, A. 1997, MNRAS, 288, 1060

\bibitem[Bodenheimer et al.(2000)]{bod00} Bodenheimer, P., Burkert, A., Klein, R.~I., \& Boss, A.~P. 2000,
in Protostars and Planets IV, ed. V. Mannings, A.~P. Boss, \&
S.~R. Russell (Tucson: Univ. Arizona Press), 675

\bibitem[Boss et al.(2000)]{boss00} Boss, A.~P., Fisher, R.~T., Klein, R.~I., \& McKee, C.~F. 2000, \apj, 528, 325

\bibitem[Chen et al.(2010)]{chen10} Chen, X., Arce, H.~G., \& Zhang, Q., et al. 2010, \apj, 715, 1344

\bibitem[Chen et al.(2008)]{chen08} Chen, X., Launhardt, R., \& Bourke, T. L., Henning, Th., \& Barnes, P. J. 2008, \apj, 683, 862

\bibitem[Chen et al.(2007)]{chen07} Chen, X., Launhardt, R., \& Henning, Th. 2007, \apj, 669, 1058

\bibitem[Chen et al.(2009)]{chen09} Chen, X., Launhardt, R., \& Henning, Th. 2009, \apj, 691, 1729

\bibitem[Choi et al.(2008)]{choi08} Choi, M., Hamaguchi, K., Lee, J.-E., \& Tatematsu, K. 2008, \apj, 687, 406

\bibitem[Evans et al.(2009)]{evans09} Evans II, N.~J., Dunham, M.~M., J{\o}rgensen, J.~K. et al. 2009, ApJS, 181, 321

\bibitem[Forbrich \& Preibisch(2007)]{forbrich07} Forbrich, J., \& Preibisch, T., 2007, \aap, 475, 959

\bibitem[Goodwin et al.(2007)]{goodwin07} Goodwin, S., Kroupa, P., Goodman, A., \& Burkert A. 2007, in Protostars and Planets V, 
ed. B. Reipurth, D. Jewitt, \& K. Keil (Tucson: Univ. Arizona Press), 133

\bibitem[Groppi et al. (2007)]{groppi07} Groppi, C.~E., Hunter, T.~R., Blundell, R., \& Sandell, G. 2007, \apj, 670, 489

\bibitem[Groppi et al. (2004)]{groppi04} Groppi, C.~E., Kulesa, C., Walker, C., \& Martin, C.~L. 2004, \apj, 612, 946

\bibitem[Harju et al.(1993)]{harju93} Harju, J. ,et al. 1993, \aap, 278, 569

\bibitem[Ho et al.(2004)]{sma04} Ho, P.~T.~P., Moran, J.~M., \& Lo, K.~Y. 2004, \apj, 616, L1

\bibitem[Knude \& Hog(1998)]{knude98} Knude, J., \& H{\o}g, E. 1998, \aap, 338, 897

\bibitem[Launhardt(2004)]{ralf04} Launhardt, R. 2004, in IAU Symp. 221, Star Formation at High Angular Resolution,
ed. M.~G. Burton, R. Jayawardhana, \& T.~L. Bourke (San Francisco: ASP), 213

\bibitem[Launhardt \& Henning(1997)]{ralf97a} Launhardt, R., \& Henning, Th. 1997, \aap, 326, 329

\bibitem[Looney et al.(2000)]{looney00} Looney, L.~W., Mundy, L.~G., \& Welch, W.~J. 2000, \apj, 529, 477

\bibitem[Machida et al. (2005)]{mach05} Machida, M.~N., Matsumoto, T., Hanawa, T., \& Tomisaka, K. 2005, MNRAS, 362, 382

\bibitem[Machida et al. (2008)]{mach08} Machida, M.~N., Tomisaka, K., Matsumoto, T., \& Inutsuka, S.-I. 2008, \apj, 677, 327

\bibitem[Masunaga et al.(1998)]{masunaga98} Masunaga, H., Miyama, S.~M., \& Inutsuka, S.-I. 1998, \apj, 495, 346

\bibitem[Maury et al.(2010)]{Maury10} Maury, A.~J., Andr\'{e}, Ph., \& Hennebelle, P., et al. 2010, \aap, 512, 40

\bibitem[Neuhauser \& Forbrich(2008)]{bally0} Neuh$\ddot{\rm a}$user, R., \& Forbrich, J. 2008, in Handbook of Star Forming
Regions, Volume II: The Southern Sky, ASP Monograph Publications,
Vol. 5. ed, Reipurth, B., 735

\bibitem[Nutter et al.(2005)]{nutter05} Nutter, D.~J., Ward-Thompson, D., \& Andr\'{e}, P. 2005, MNRAS, 357, 975

\bibitem[Ohashi et al.(1997)]{ohashi97} Ohashi, N., Hayashi, M., Ho, P.~T.~P., \& Momose, M. 1997, \apj, 475, 211

\bibitem[Ossenkopf \& Henning(1994)]{ossenkopf94} Ossenkopf, V., \& Henning, Th. 1994, \aap, 291, 943

\bibitem[Qi(2005)]{sma05} Qi, C. 2005, MIR Cookbook (Cambridge: Harvard), http://cfa-www.harvard.edu/~cqi/mircook.html

\bibitem[Sault et al.(1995)]{sault95} Sault, R.~J., Teuben, P.~J.,
\& Wright, M.~C.~H. 1995, in ASP Conf. Ser. 77, Astronomical Data
Analysis Software and Systems IV, ed. R.~A. Shaw, H.~E. Payne, \&
J.~J.~E. Hayes (San Francisco: ASP), 433

\bibitem[Teixeira et al.(2007)]{teixeira07} Teixeira, P.~S., Zapata, L.~A., \& Lada, C.~J. 2007, \apj, 667, L179

\bibitem[Tohline(2002)]{tohline02} Tohline, J.~E. 2002, ARA\&A, 40, 349

\bibitem[Truelove et al.(1997)]{truelove97} Truelove, J.~K., Klein, R.~I., \& McKee, C.~F. et al. 1997, \apj, 489, L179

\bibitem[Truelove et al.(1998)]{truelove98} Truelove, J.~K., Klein, R.~I., \& McKee, C.~F. et al. 1998, \apj, 495, 821

\bibitem[Tsuribe \& Inutsuka(1999)]{tsuribe99} Tsuribe, T., \& Inutsuka, S. 1999, \apj, 523, L155

\bibitem[van den Ancker(1999)]{ancker99} van~den~Ancker, M. 1999, Ph.D. thesis, Univ. Amsterdam

\bibitem[Volgenau et al.(2006)]{vol06} Volgenau, N.~H., Mundy, L.~G., Looney, L.~W., \& Welch, W.~J. 2006, \apj, 651, 301

\bibitem[Wilking et al.(1997)]{wilking97} Wilking, B.~A., McCaughrean, M.~J., \& Burton, M.~G., et al. 1997, \aj, 114, 2029

\end{thebibliography}
\end{document}